# Activated decay of a metastable state: Transient times for small and large dissipation

**M. V. Chushnyakova[1], I. I. Gontchar[2], N. A. Khmyrova[2]**

[1]Physics Department, Omsk State Technical University, Omsk 644050, Russia
[2]Physics and Chemistry Department Omsk State Transport University, Omsk 644046, Russia

Corresponding author:
e-mail: maria.chushnyakova@gmail.com

ORCID:
0000-0003-0891-3149 (MVC)
0000-0002-9306-6441 (IIG)

**Abstract**
The time evolution of the thermally activated decay rates is considered. This evolution is of particular importance for the recent nanoscale experiments discussed in the literature, where the potential barrier is relatively low (or the temperature is relatively high). The single-molecule pulling is one example of such experiments. The decay process is modeled in the present work through computer solving the stochastic (Langevin) equations. Altogether about a hundred of high precision rates have been obtained and analyzed. The rates are registered at the absorption point located far beyond the barrier to exclude the influence of the backscattering on the value of the quasistationary decay rate. The transient time, i.e. the time lapse during which the rate attains half of its quasistationary value, has been extracted. The dependence of the transient times upon a damping parameter is compared with that of the inverse quasistationary decay rate. Two analytical formulas approximating the time-dependences of the numerical rates are proposed and analyzed.

## 1. Introduction

We present the results of numerical modeling of the one-dimensional motion of a Brownian particle escaping from a potential pocket over a barrier. This pocket is illustrated in Fig. 1. Initially, the Brownian particles are at rest at the metastable state. Due to the action of thermal fluctuations, a flux over the barrier (and then over the absorption (sink) point $q_s$) appears, increases, and becomes quasistationary (see Fig. 2 below). This model is widely employed in the literature for describing different processes like nuclear fission [1–3], single-molecule pulling [4,5], cracks formation [6], dissociation of a molecule [4,7], chemical reactions [7–9], survival of a contact formed by the adhesive interaction [10], etc.

    In our former articles, we have studied in detail (mostly through computer modeling) the value of the quasistationary decay rate, $R_D$, considering its dependence upon the shape of the barrier [11], the influence of an extra degree of freedom [12], and the behavior of $R_D$ in the case of extremely weak friction [13,14]. We also have considered the influence of the backscattering [15] as well as the case of high temperature (low barrier) [16]. When possible, we compared our quasistationary numerical rates with the approximate analytical ones obtained using the famous Kramers approach [7]. The work by others is also mostly focused on the quasistationary stage of the decay process (see e.g. reviews [17,18] and references therein).

    In the present paper, we are looking for some regularities in the time evolution of the rate which is beyond the scope of the Kramers treatise [7]. We are aware only about three publications devoted to this subject [19–21].

    However, the importance of the rate evolution, as it reaches its quasistationary value, increases as the barrier becomes lower, i.e. the governing parameter

$$G = \frac{U_b}{\theta} \qquad (1)$$

becomes smaller. Here $U_b$ denotes the height of the potential barrier; $\theta$ is the average thermal energy of the particle motion near the bottom of the pocket.

What are the typical values of the transient times? What is the shape of the time-dependent rate at the transient stage and is it possible to find an analytical profile for it? These are the questions we are trying to answer in the present work.

## 2. Numerical modeling

When modeling any kind of a physical process, the potential energy plays a crucial role. In the present work, we use the potential energy constructed of two smoothly jointed pieces of parabola:

$$U(q) = \begin{cases} \dfrac{C_c(q - q_c)^2}{2} & \text{at } q < q_m; \\ U_b - \dfrac{C_c(q - q_b)^2}{2} & \text{at } q_m \leq q. \end{cases} \quad (2)$$

Here the subscript "$c$" indicates the bottom of the pocket, "$b$" refers to the barrier top, $C_c$ denotes the stiffness. The input parameters for the potential are $U_b = 6$ (unless not specified otherwise), $q_c = 1.0$, $q_b = 1.6$. The matching point $q_m$ reads:

$$q_m = \frac{q_c + q_b}{2}. \quad (3)$$

The physical meaning of the generalized coordinate $q$ depends upon the physical problem under consideration. For instance, in the case of nuclear fission, it is proportional to the distance between the nascent fragments of the fissioning atomic nucleus. In the case of molecule dissociation, the principle coordinate is the distance between the forming ions or other parts of the molecule. When they consider the adhesive contact, the generalized coordinate might be proportional to the radius of the contact area. To avoid the binding to the specific problem and to make the results of our research valuable for a wider audience, we prefer to use the dimensionless coordinate.

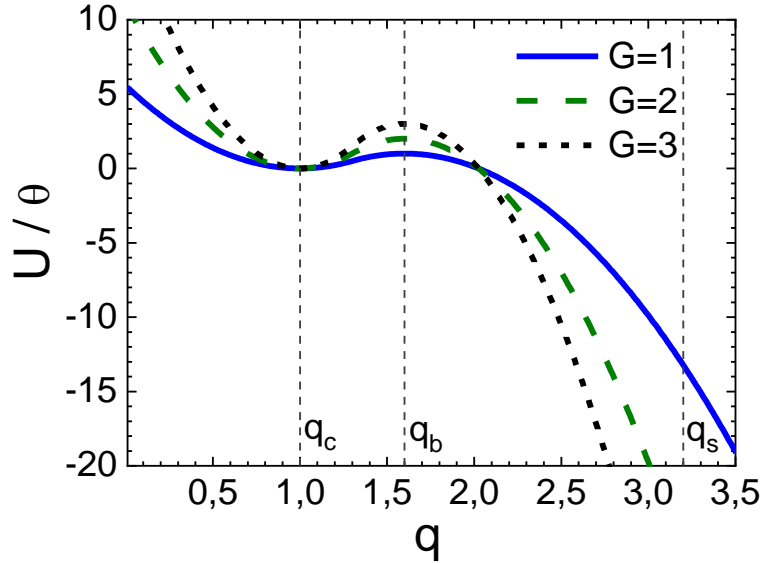

**Fig. 1.** Typical dimensionless potentials with a metastable well (left, $q_c$), barrier (middle, $q_b$), and sink point (right, $q_s$). $\theta$ denotes the thermal energy. For the definition of $G$ see Eq. (1)

Three potential landscapes are presented in Fig. 1 for three different values of the governing parameter. The potential energies in this figure are divided by $\theta$ since we know that it is $U_b/\theta$ that governs the process [22]. Actually, there are two dimensionless parameters controlling the decay process [22]. The second one is the damping parameter

$$\varphi = \frac{\eta}{m\omega}. \quad (4)$$

Here $\eta$ is the friction parameter, $m$ denotes the inertia parameter, $\omega$ stands for the frequency of the particle oscillations near the bottom of the pocket.

There are three ways for stochastic modeling of the decay process using the Langevin type equations:
- the stochastic ordinary differential equations in the phase space (two Langevin equations for the coordinate $q$ and conjugate momentum $p$, LEqp) [1–3,12,23];
- the stochastic ordinary differential equation in the configuration space (single Langevin equation for the coordinate $q$, LEq) [11,24–27];
- the stochastic ordinary differential equation for the action $I$ (or energy $E$) variable (single Langevin equation for the action, LEI) [14].

We have some expertise in all these approaches from our previous studies therefore we know that the applicability domains of these approaches are quite different. The LEqp produce correct results for any value of the damping parameter but often take extremely long computer time. The LEq takes much less computer time but results in the correct rates only at $\varphi \gg 1$.

To estimate the accuracy of the LEq, let us consider the approximate Kramers formula for $R_D$, $R_{Kqp}$:

$$R_{Kqp} = \left[\left(\frac{\varphi^2}{4}+1\right)^{1/2} - \frac{\varphi}{2}\right]\frac{\omega}{2\pi}\exp(-G). \quad (5)$$

Expanding Eq. (5) into the power series with respect to the small parameter $4\varphi^{-2}$ results in

$$R_{Kqp} \approx \left(1 - \frac{1}{\varphi^2}\right)\frac{\omega}{2\pi\varphi}\exp(-G). \quad (6)$$

Concerning the quasistationary rate, applying the LEq is equivalent to using

$$R_{Kq} = \frac{\omega}{2\pi\varphi}\exp(-G). \quad (7)$$

Thus, the error of the LEq approach in comparison to the LEqp approach is of the order of $\varphi^{-2}$, i.e. about 1% at $\varphi = 10$.

The rates obtained within the framework of the LEI are close to the correct ones at $\varphi \ll 1$, but at $\varphi = 0.01$ the error of $R_D$ resulting from the LEI can easily reach 20% in comparison to $R_D$ resulting from the LEqp [14]. That is why in the present work we use for the modeling either the LEqp at $\varphi < 4$ and the LEq otherwise.

The algorithm of numerical modeling within the framework of the LEqp is described in detail in [15,16,20] therefore we omit it here. Analogously, one finds a comprehensive description of the LEq method (the reduced Langevin equation) in Refs. [11,24,25] and we do not discuss it here too. In the present modeling, within the framework of both these algorithms, initially (i.e. at $t = 0$) the Brownian particle is positioned at the bottom of the potential well ($q = q_c, p = 0$). Each run of the modeling (solving numerically the relevant stochastic differential equations) results in a sequence of $N_{tot}$ trajectories (typically $10^5 \div 10^6$). Each trajectory is terminated not later than at $t = t_D$. Some of the particles (trajectories) arrive at the absorptive border (sink) $q_s = 3.2$ before $t_D$. This rather large value of $q_s$ excludes its influence on the resulting quasistationary rate due to the backscattering (see [12,15] for details). The numerical decay rate at any moment of time is evaluated within this approach in the following manner:

$$R_{sn}(t) = \frac{1}{N_{tot} - N_s(t)}\frac{\Delta N_s}{\Delta t}. \quad (8)$$

Here $N_s(t)$ is the number of Brownian particles which have arrived at $q_s$ by the time moment $t$; $\Delta N_s$ is the number of the particles which have reached the sink during the time lapse $\Delta t$.

### 3. Results and discussion

A typical rate resulting from the modeling is displayed in Fig. 2. It reaches a quasistationary value $R_D$ (horizontal line) after some transient stage. We characterize this stage by the time interval $T_{sn}$ (transient time) during which the rate attains the value equal to $0.5R_D$.

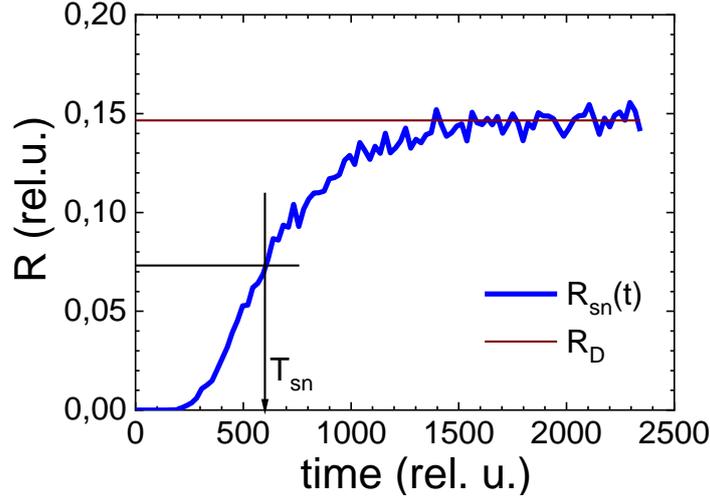

**Fig. 2.** Time evolution of the decay rate.

More rates are presented in Fig. 3. The rates shown by the lines with (without) symbols are calculated using $U_b = 6$ ($U_b = 8$), respectively. The values of $\theta$ were adjusted accordingly to keep the governing parameter $G = 2.4$ fixed. Also, the values of the stiffness $C_c$ (see Eq. (2)) and frequency $\omega$ become different; that is why in Fig. 3 the rates and time are scaled using $\omega$. In each panel, two different values of the friction parameter $\eta$ are used keeping, however, the damping parameter $\varphi$ unchanged for two runs. Comparing the rates in each panel of Fig. 3 one concludes that not only the values of $R_D$ but the whole time-dependent rate is completely defined by the parameters $G$ and $\varphi$. Thus, the transient time $T_{sn}$ must not be an exception.

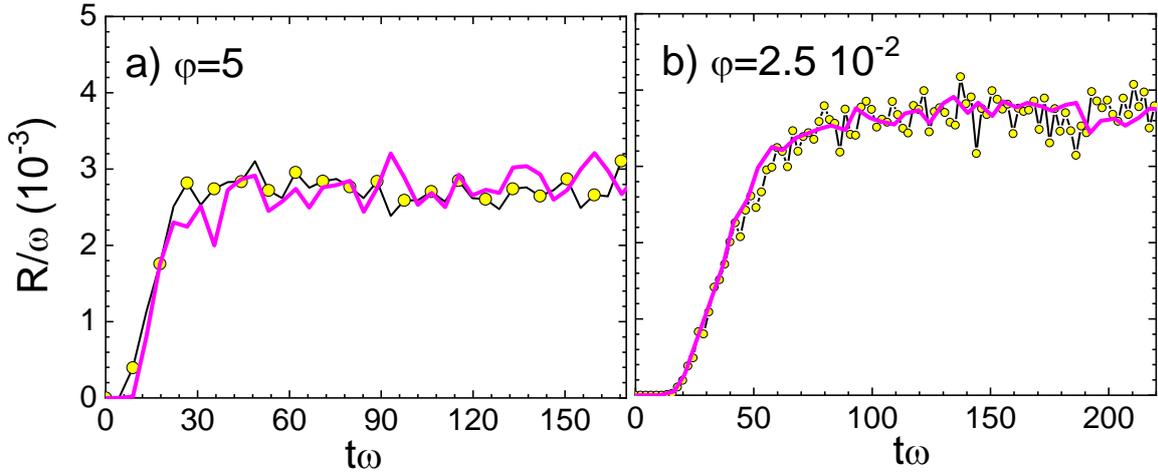

**Fig. 3.** For two values of the damping parameter, the time-dependent rates calculated with $U_b = 6$ and 8 are shown by the curves with and without symbols, respectively. $G = 2.4$.

In Fig. 4a we show some values of $T_{sn}\omega$ versus the damping parameter for $G =0.7, 1.5, 2.4, 3.5$, and 5.0. Clearly, the values of $T_{sn}(\varphi)$ corresponding to the different values of $G$ are sometimes rather close. In order to extract this transient time numerically with the proper accuracy, we need, on the one hand, to calculate $R_{sn}(t)$ [see Eq. (8)] with small enough time lapse $\Delta t$ and, on the other hand, the time interval under consideration should inevitably contain the target value of $T_{sn}$. Thus, first of all, we have to choose an optimal upper limit of time, $t_{up}$. We have found that the dependence

$$t_{up}\omega = 9\varphi^{-1} + 16\varphi^{-1/2} + 32\varphi \qquad (9)$$

works well for all values of the governing parameter. This dependence is shown in Fig. 4a by the solid line without symbols.

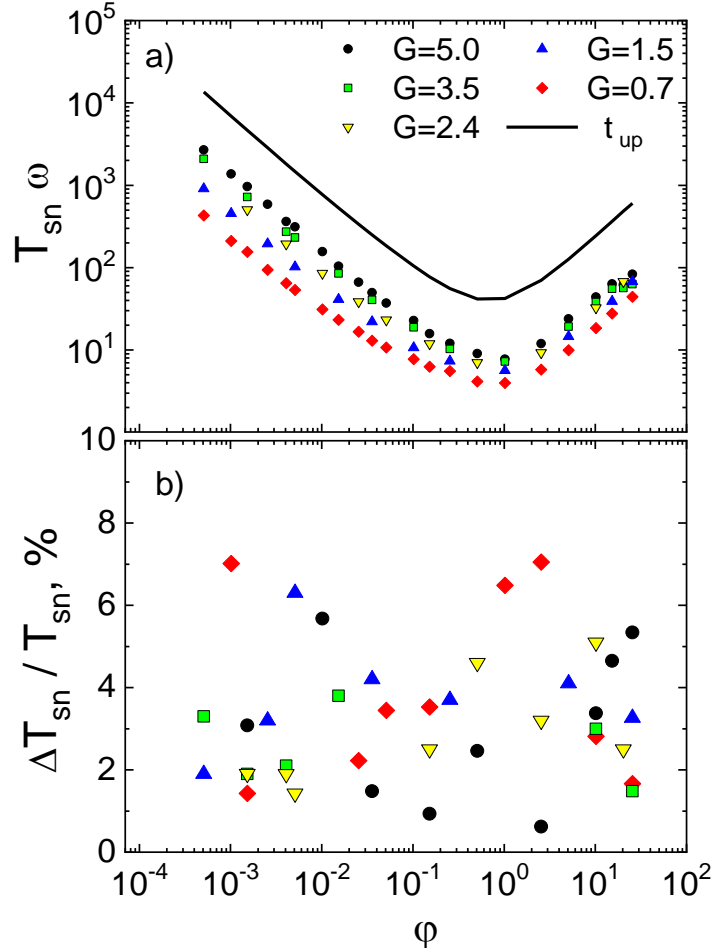

**Fig. 4.** a) $T_{sn}\omega$ (symbols) and the time lapse $t_{up}\omega$ defined by Eq. (9) (solid line) versus $\varphi$;
b) the relative uncertainty of the transient time versus $\varphi$.

Of course, the numerical transient times extracted from the results of modeling are subject of errors. We estimate these errors (uncertainties) in the following manner. The set of the trajectories corresponding to the given values of $\varphi$ and $G$ is processed thrice: i) with $t_{up}$, ii) with $0.8 t_{up}$, and iii) $0.7 t_{up}$. The maximum difference between the three values of $T_{sn}$ is taken as the uncertainty of it, $\Delta T_{sn}$. The relative uncertainties obtained according to this algorithm are presented in Fig. 4b. One sees that the routine of extracting $T_{sn}$ from the raw data is quite accurate: the relative uncertainty never exceeds 10%.

In Fig. 5 the dependences of $T_{sn}\omega$ upon the damping parameter evaluated for $G =$ 0.7, 1.5, 2.4, 3.5, 5.0 are presented in more detail. Since the values of $T_{sn}(\varphi)$ corresponding to different values of $G$ are rather close (although still different), we multiply each curve by a proper factor to make the figure readable.

The shape of the $T_{sn}(\varphi)$-dependence resembles the inverse $\varphi$-dependence of the $R_D$-values (see, e.g., Fig. 5 in Ref. [28] and Fig. 1 in Ref. [29]). Therefore, it seems useful to check whether the dependences $T_{sn}(\varphi)$ and $R_D^{-1}(\varphi)$ really coincide. We perform this comparison in Fig. 6. Clearly, there is no exact coincidence although for $G = 0.7$ and $\varphi < 0.1$ it is possible to make the curves overlapping.

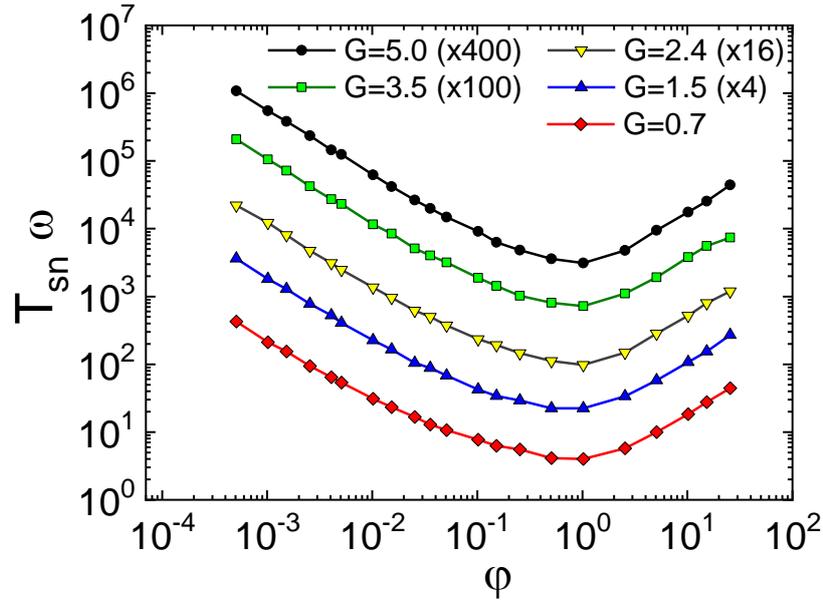

**Fig. 5.** $T_{sn}\omega$ versus $\varphi$ for different values of $G$. To make the figure more readable, for some values of $G$ the transient times are multiplied by the proper factors shown in the figure

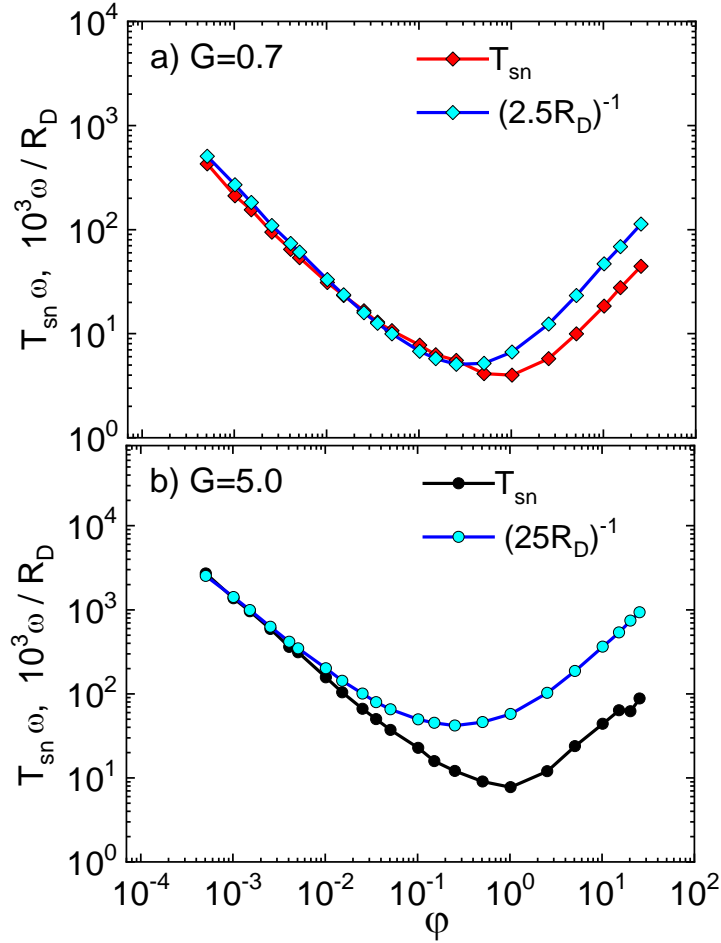

**Fig. 6.** The $\varphi$-dependence of the transient time $T_{sn}$ is compared with that of the inverse quasistationary rate for two values of the governing parameter. The multipliers 2.5 (a) and 25 (b) for $R_D$ are included for convenience.

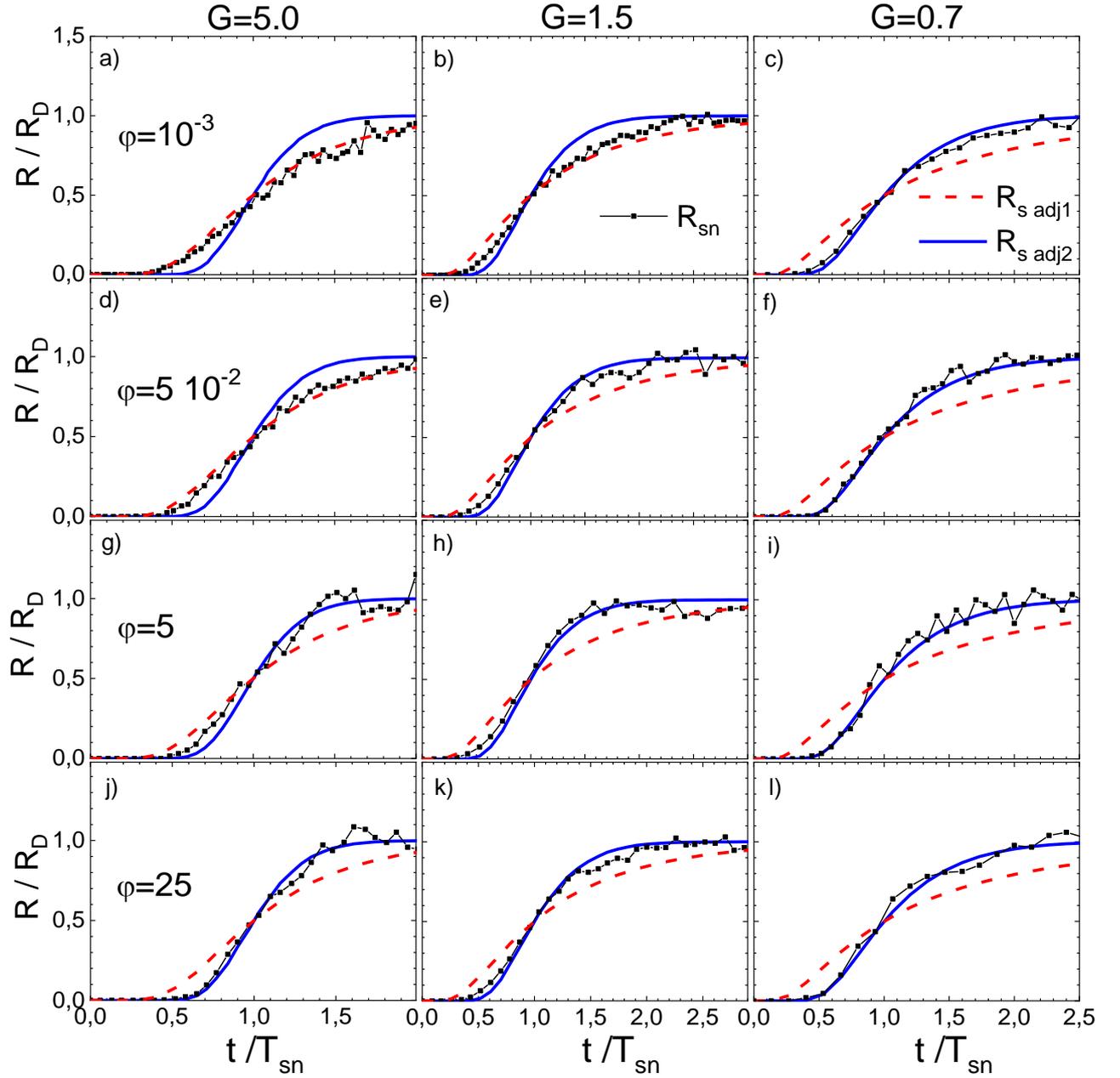

**Fig. 7.** Time-dependent rates (over their quasistationary values) as functions of time (over $T_{sn}$) for three values of the governing parameter and four values of the damping parameter indicated in the figure. The curves with symbols correspond to $R_{sn}(t)$, red dashed lines denote $R_{s\,adj1}(t)$, blue solid lines are for $R_{s\,adj2}(t)$.

As the next step of our study, we try to find an analytical formula for the $R_{sn}(t)$-dependence. As the starting point, we use an ansatz similar to the approximate result from Ref. [19]

$$R_{s\,adj1}(t) = R_D \exp\left\{\frac{-G}{\exp(t/\tau_{m1}) - 1}\right\}, \tag{10}$$

where

$$\tau_{m1} = \frac{T_{sn}}{\ln(1 + G/\ln 2)}. \tag{11}$$

In Fig. 7 the adjusted approximate rates $R_{s\,adj1}(t)$ (red dashed lines without symbols) are compared with the numerical ones $R_{sn}(t)$ (broken black lines with symbols). Examining Fig. 7 one sees that sometimes the $R_{s\,adj1}(t)$ provides a very good approximation for the numerical rate (panels a, d) but mostly it does not. In most of the panels, the numerical rate increases steeper than $R_{s\,adj1}(t)$ does. Therefore, we try to approximate the rate using

$$R_{s\,adj2}(t) = R_D \exp\left\{\frac{-G}{\exp[(t/\tau_{m2})^2] - 1}\right\}, \tag{12}$$

where

$$\tau_{m2} = \frac{T_{sn}}{[\ln(1 + G/\ln 2)]^{1/2}}. \tag{13}$$

The rates $R_{s\,adj2}(t)$ (blue solid lines without symbols) result in a better approximation for $R_{sn}(t)$ for many more cases presented in Fig. 7. However, one should be careful with tempting conclusions due to the limited number of examples presented in this figure. What can be stated that there is no one case where the rate $R_{sn}(t)$ rises significantly steeper than the $R_{s\,adj2}(t)$.

We believe, the next step of our work should be finding an exponent (apparently between 1 and 2) in an approximate formula

$$R_{s\,adj\,u}(t) = R_D \exp\left\{\frac{-G}{\exp[(t/\tau_{mu})^u] - 1}\right\}, \tag{14}$$

providing the best fit for the transient stage of each rate. This program is presently underway.

**4. Conclusions**
In the present work, the time evolution of the thermally activated decay rates is considered. This might be timely for the recent nanoscale experiments with relatively low barriers. The decay process is modeled solving the stochastic (Langevin) equations numerically. Altogether about 100 high precision rates have been obtained and for each rate the transient time $T_{sn}$ (i.e. the time lapse during which the rate reaches half of its quasistationary value) has been recovered. The dependence of the $T_{sn}$ upon the dimensionless damping parameter $\varphi$ has been established and compared with the $\varphi$-dependence of the inversed quasistationary decay rate.

Simple analytical formulas (10), (12) have been tested for the adjustment of the $R_{sn}(t)$-dependence: in some cases, Eq. (10) works better, in other cases Eq. (12) does so. To find the universal analytical formula for the time-dependent decay rate at any values of the governing and damping parameters, we propose to use an analytical formula (14) similar to Eqs. (10) and (12) with the adjustable exponent $u$. This exponent is expected to take a value between 1 and 2. Finding its value providing the best fit for the transient stage of each rate is in progress.